\documentclass[a4paper]{jpconf}
\usepackage{graphicx}
\begin{document}
\title{Spectral properties of Compton inverse radiation: Application of Compton beams}

\author{Eugene Bulyak$^1$ and Junji Urakawa$^2$}

\address{$^1$ NSC KIPT, Kharkov, Ukraine}
\address{$^2$ KEK, Tsukuba, Ibaraki, Japan}

\ead{bulyak@kipt.kharkov.ua}

\begin{abstract}
Compton inverse radiation emitted due to backscattering of laser pulses off the relativistic electrons possesses high spectral density and high energy of photons -- in hard x--ray up to gamma--ray energies -- because of short wavelength of laser radiation as compared with the classical electromagnetic devices such as undulators.

In this report, the possibility of such radiation to monochromatization by means of collimation  is studied. Two approaches have been considered for the description of the spectral--angular density of Compton radiation based on  the classical field theory and on the quantum electrodynamics. As is shown, both descriptions produce similar total spectra.
On the contrary, angular distribution of the radiation is  different: the classical approach predicted a more narrow radiation cone.
Also proposed and estimated is a method of the `electronic' monochromatization based on the electronic subtraction of the two images produced by the electron beams with slightly different energies. A `proof--of--principle' experiment of this method is proposed for the LUXC facility of KEK (Japan).
\end{abstract}

\section{Introduction}
Sources of Compton radiation, in which photons of intense laser pulse scattered off from relativistic electrons, are able to produce bright x--ray beams with narrow bandwidth.
The process of Compton scattering can be treated as two--particle elastic scattering: one of the particle is represented by the electron, another one by the photon. Since the laser photon has negligibly small energy as compared with electron's, the recoil of electron is negligibly small -- the photon is scattered off within a narrow cone along electron's trajectory.

Such sources have a substantial potential for applications in different areas of medicine, biology, physics, etc. This potential is emerged not only from its brightness, but tunability and `quasimonochromaticity' of the spectrum.

The report highlights spectral properties of Compton sources and presents a potential application of these sources for x--ray angiography, which possesses a substantial advantage over the conventional methods.

\section{Spectrum of collimated Compton inverse radiation}
The spectrum of inverse Compton radiation is determined by two relations, (i) kinematic dependence of the energy of a scattered off quantum, $E_\mathrm{x}$ upon the crossing angle $\phi$ between the laser photon and the electron, energy of both the electron and the photon, and the scattering angle $\psi$, and (ii) the differential cross section -- dependence of probability of scattering the photon at angle $\psi$. Within the small--angle approximation,  $\psi\ll 1$ (the crossing angle $\phi = 0$ corresponds to the head--on collision), neglected the electron recoil, these relations read (\cite{buskoepac04}):
\[
E_\mathrm{x}=\frac{2\gamma^2(1+\cos\phi )E_\mathrm{las}}{1+\gamma^2\psi^2}\;; \qquad
\mathrm{d}\sigma= 8\pi r^2_0\frac{\psi\gamma^2
\left(1+\gamma^4\psi^4\right)}{\left(1+\gamma^2\psi^2\right)^4}
\mathrm{d}\psi \; ,
\]
where $\gamma = E_\mathrm{e}/m_\mathrm{e}c^2$ is the Lorentz factor of the electron, $r_0$ is the classical electron radius.

\begin{figure}[h]
\begin{minipage}{0.4\textwidth}
\center
\includegraphics[width=0.8\textwidth]{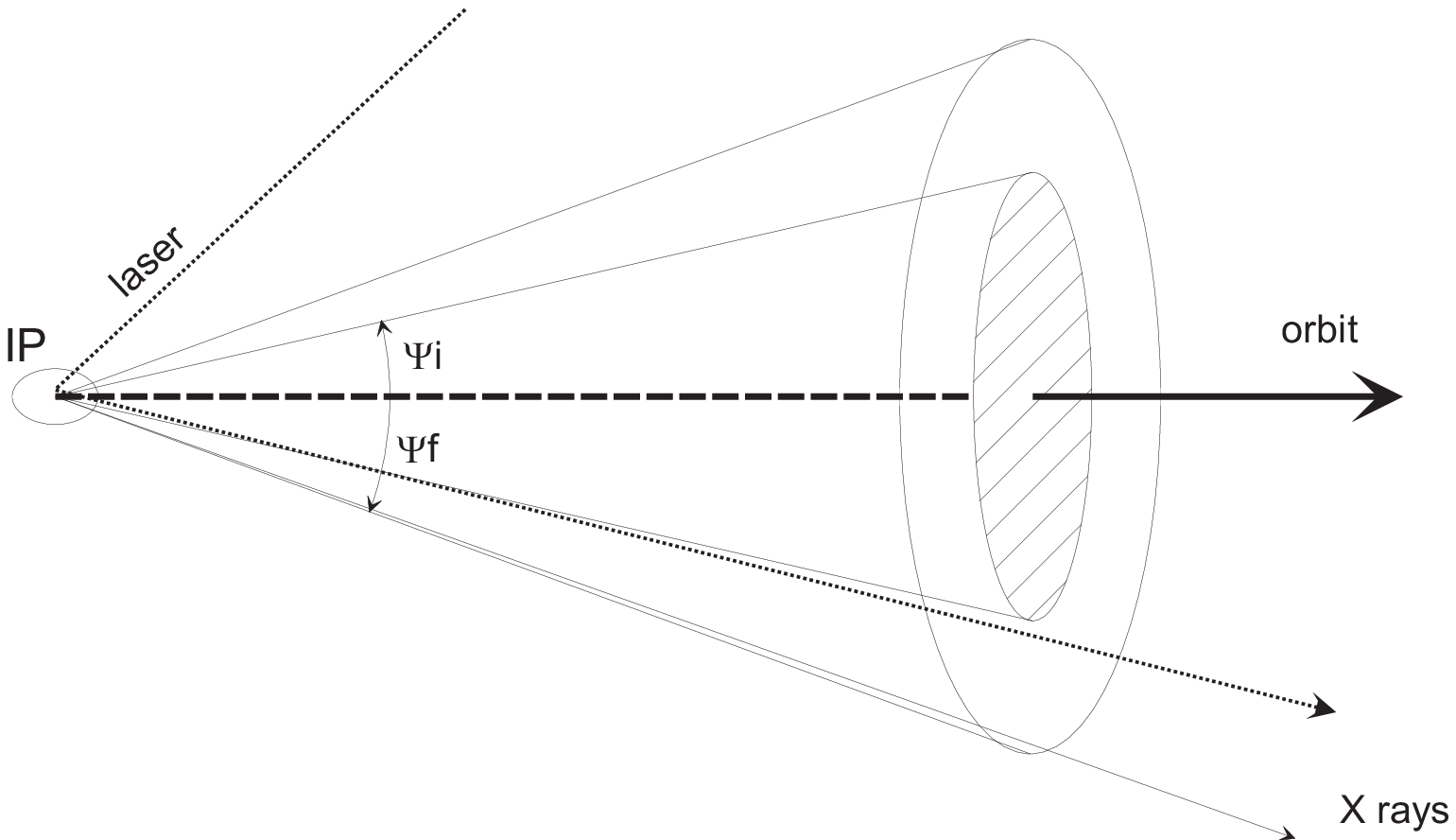}
\caption{\label{fig1a}Scheme of the Compton radiation collimation.}
\end{minipage}\hspace{2pc}
\begin{minipage}{0.4\textwidth}
\includegraphics[width=0.8\textwidth]{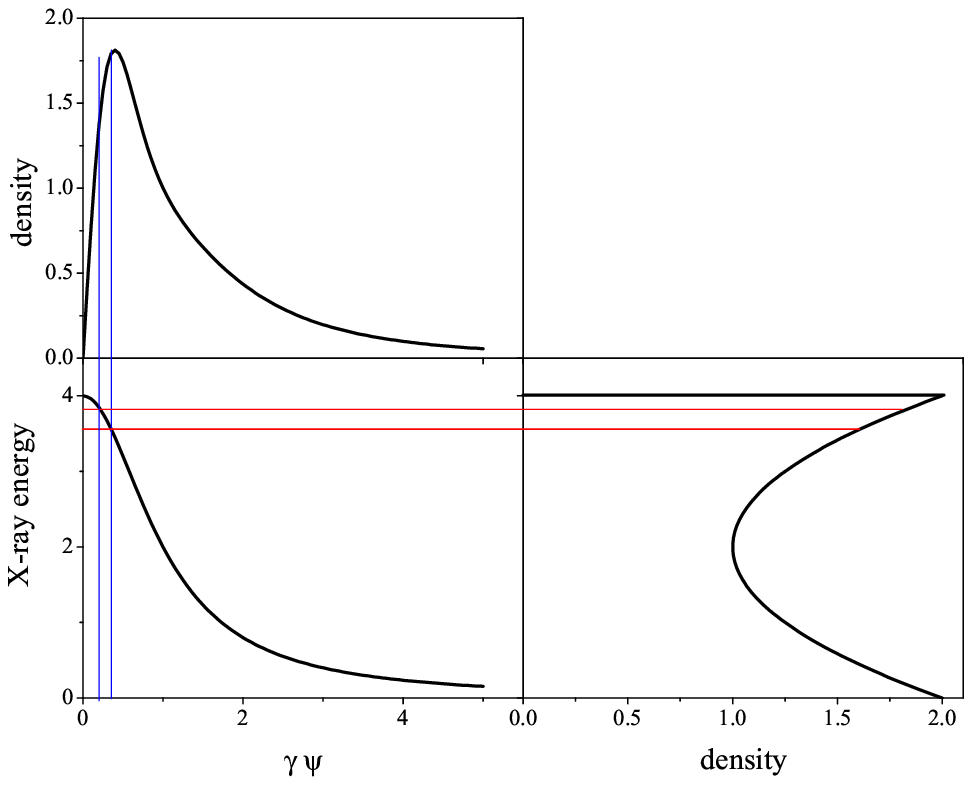}
\caption{\label{fig1b}Spectral-angular density of the Compton radiation.}
\end{minipage}
\end{figure}

Convolution of the cross section with distribution functions of both the electron bunch and the photon pulse taking into account the kinematic relation will produce a real collimated spectrum of the radiation.
As it follows from the kinematics, the width of spectrum is determined by the collimation angles $\psi_i\le \psi \le \psi_f$, as is depicted in Fig.\ref{fig1a}. The spectral--angular density of the collimated Compton radiation for an ideal case (monoenergetic electrons and laser photons with parallel trajectories) is presented in Fig.\ref{fig1b}.

For the vast majority of practical sources the inner collimation angle is zero, $\psi_i = 0$. We will consider below this particular case of an iris collimator.

In general, the spectrum is dependent on the following factors: the collimator opening angle, the energy spread of electrons and photons and the angular spread of their trajectories at the interaction point (IP). In practical cases, the angular spread of photon `trajectories' produces least impact on the spectrum, as it can be seen from the kinematics: Its effect is proportional to the spread squared because IP is usually set up at the laser waist, while the electron spread impact is $\gamma^2$ times larger. Also the spread of electron energy within the bunch usually much higher than that of photons. Therefore below we will consider effects of the electron bunch phase volume -- spreads of energy and trajectories -- upon the collimated spectrum of scattered off laser quanta.

\subsection{Spread of electrons energy}
\begin{figure}[h]
\begin{minipage}{0.45\textwidth}
\includegraphics[width=\textwidth]{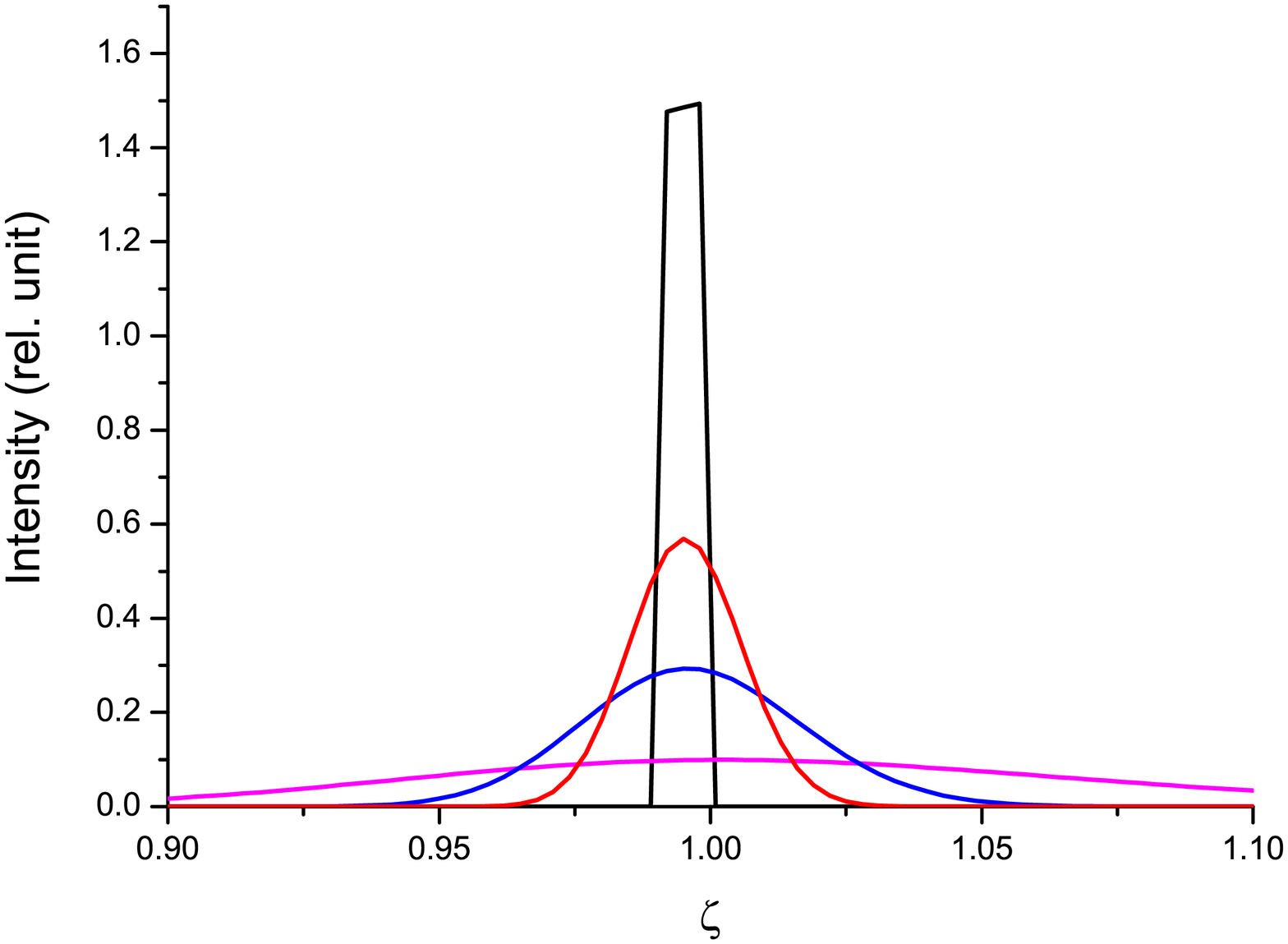}
\caption{\label{fig2a} X-ray spectra for the electron energy spreads $\sigma_e\gamma = (0,
0.01, 0.02, 0.03)$ into collimating range $\psi\gamma=(0,0.1)$.}
\end{minipage}\hspace{2pc}%
\begin{minipage}{0.45\textwidth}
\includegraphics[width=0.9\textwidth]{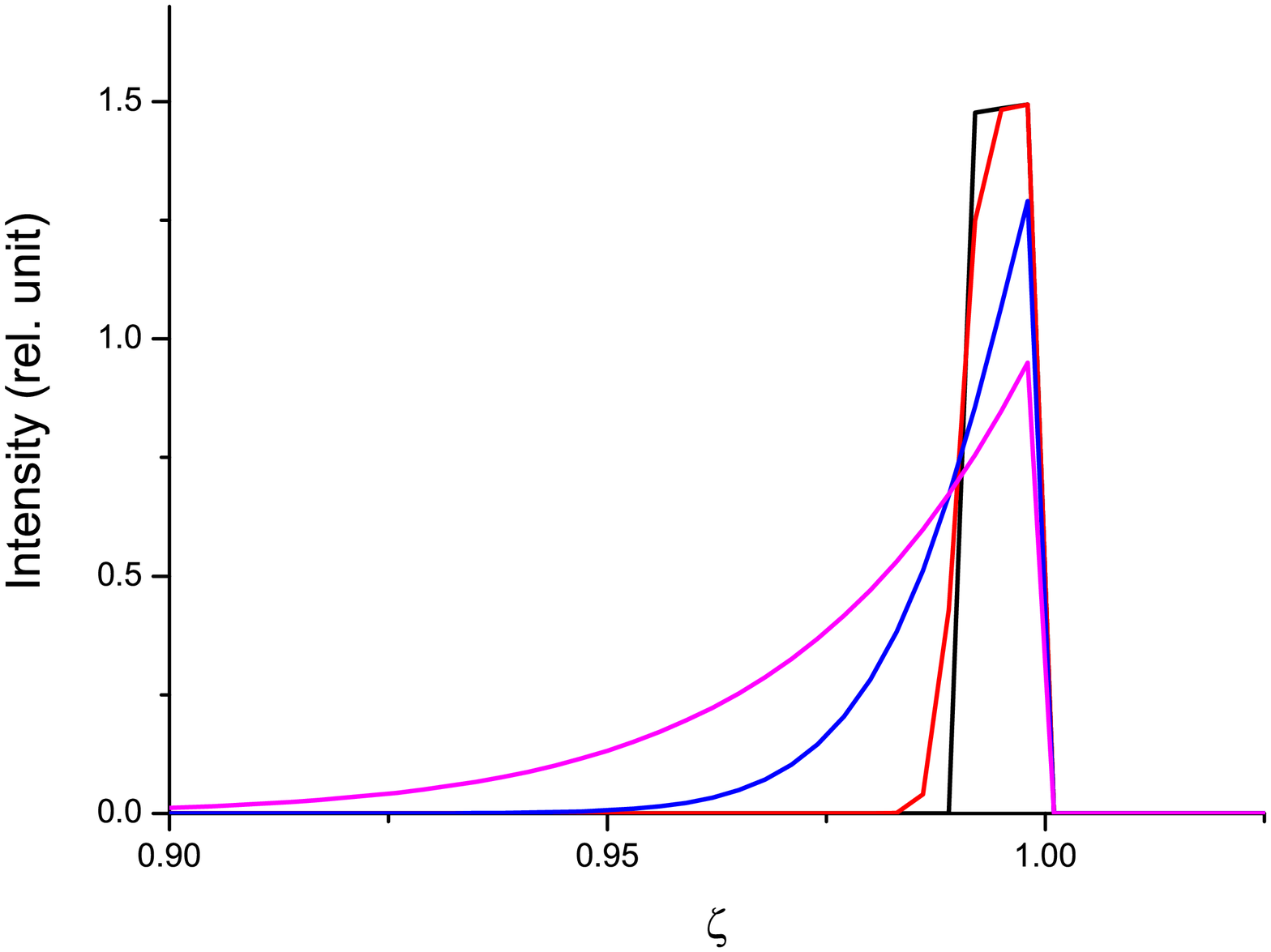}
\caption{\label{fig2b} X-ray energy spectra for the electron trajectories
spreads $\sigma_e\gamma =(0.01, 0.1, 0.5, 1.0)$ into collimating range $\psi\gamma=(0,0.1)$.}
\end{minipage}
\end{figure}

The  energy spread dilutes the spectral--angular density curve in Fig.\ref{fig1b} along the energy axis. Since the energy spread in bunches of Compton sources is usually small -- from a few percents down to fraction of a percent, -- the corresponding partial energy spread of x--rays is equal to doubled the electrons': For the normal (Gaussian) distribution of electrons' energy in the bunch, the `pin-hole' collimated x--ray beam has the doubled reduced dispersion: $\sigma_x/E_x\approx 2 \sigma_e/E_e$. (Detailed study on the spectrum for head-on collision is presented in \cite{sun09}.) Analytical collimated spectra for finite energy spread of electrons with parallel trajectories are presented in Fig.\ref{fig2a}.

\subsection{Angular spread of electrons trajectories}
The angular spread of electrons' trajectories induces much more widening into the spectrum since the radiation is emitted within narrow cone along the trajectory. (It should be noted that a quantum emitted at the angle $1/\gamma$ to the electron trajectory has a half of the energy of that emitted along the trajectory.) Also tight focusing of the bunches at the interaction point aimed to gain the yield of x rays, increases the angular spread.

A collimated spectrum with account for the angular spread has asymmetric and more complicated shape than that due to the energy spread. It is specified with steep high--energy cutoff (stemmed from the energy conservation law) and long low--energy tail, see Fig.\ref{fig2b}.

\begin{figure}[h]
\includegraphics[width=18pc]{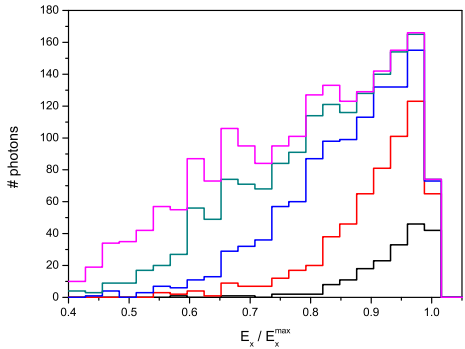}\hspace{2pc}%
\begin{minipage}[b]{15pc}\caption{\label{fig3spectra} Spectra for collimation $\psi\gamma  = 0.2,0.4,0.6,0.8,1.0$ (from the narrowest to widest profiles). Angular horizontal and vertical spreads in the bunch are $\gamma \sigma_{x,y} = 0.08,0.32$.}
\end{minipage}
\end{figure}

Examples of simulated spectra for different collimation opening angles is presented in Fig.\ref{fig3spectra} (the parameters for simulation were taken similar to those of LUCX facility, see \cite{fukuda13}).

From the study on the collimated x-ray spectra generated by Compton sources the points are followed:
\begin{itemize}
\item The spectrum width is limited by the high energy cutoff, no photons with higher energy.
\item The spectrum has the maximum close to the high-energy cutoff.
\item High-energy cutoff of the spectrum is steep.
\item The collimator opening angle  should not be more narrow than the (approximately doubled) angular spread of electron trajectories at IP. More narrow collimation will reduce the spectral maximum but not the width.
\end{itemize}

\section{Application of Compton sources for angiography}
Angiography, a medical x--ray based imaging technique to visualize the inside of blood vessels and organs of the body, uses a radio-opaque contrast agent injected into the blood vessel. The difference between x--ray images without and with the contrast agent in the blood produces a picture of the blood flow.

Iodine--based radiocontrast agents are most commonly used in angiography with maximum contrast energy of x--rays just above the K--edge of iodine, 33.17\,keV. Hence, the x--rays within a band of $\sim$30--35\,keV most effectively produce the images. Other energies are redundant and even detrimental since they cause additional radiation load.

\begin{figure}[h]
\includegraphics[width=20pc]{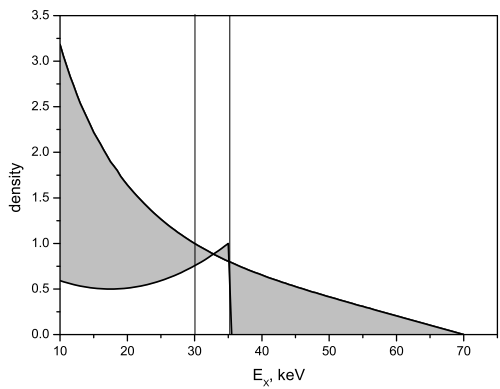}\hspace{2pc}%
\begin{minipage}[b]{14pc}\caption{\label{bremCompSpectra}Bremstrahlung (70\,keV electrons and tungsten radiator) and Compton spectra. the vertical lines limit a useful x-ray energy band, the gray areas beyond show redundant x--ray radiation load from the bremsstrahlung spectrum.}
\end{minipage}
\end{figure}

The Compton x--ray sources with the tunable energy peak possess certain advantages over the conventional bremsstrahlung sources. In Fig.\ref{bremCompSpectra}, there an ideal Compton spectrum and a bremsstrahlung one are presented that produce equal fluxes within 30--35\,keV energy range.

As it can be seen from the picture, the conventional source produces much more background x--ray quanta (the gray areas beyond 30--35\,keV range).

\subsection{Simulation of Compton angiography}
For verification of Compton sources applicability for angiography, a digital model has been created. The models simulates the image of the tissue composed from the skeletal muscle, the bone cortical, blood and iodine components (x--ray attenuation coefficients were taken from NIST \cite{nist}). Distribution of the energy of x--ray quanta is simulated with Monte Carlo method according to the collimate ideal Compton spectrum. Distribution of the impinging quanta over the tissue face is uniform, random.

\begin{figure}[h]
\centering{
\includegraphics[width=0.4\textwidth]{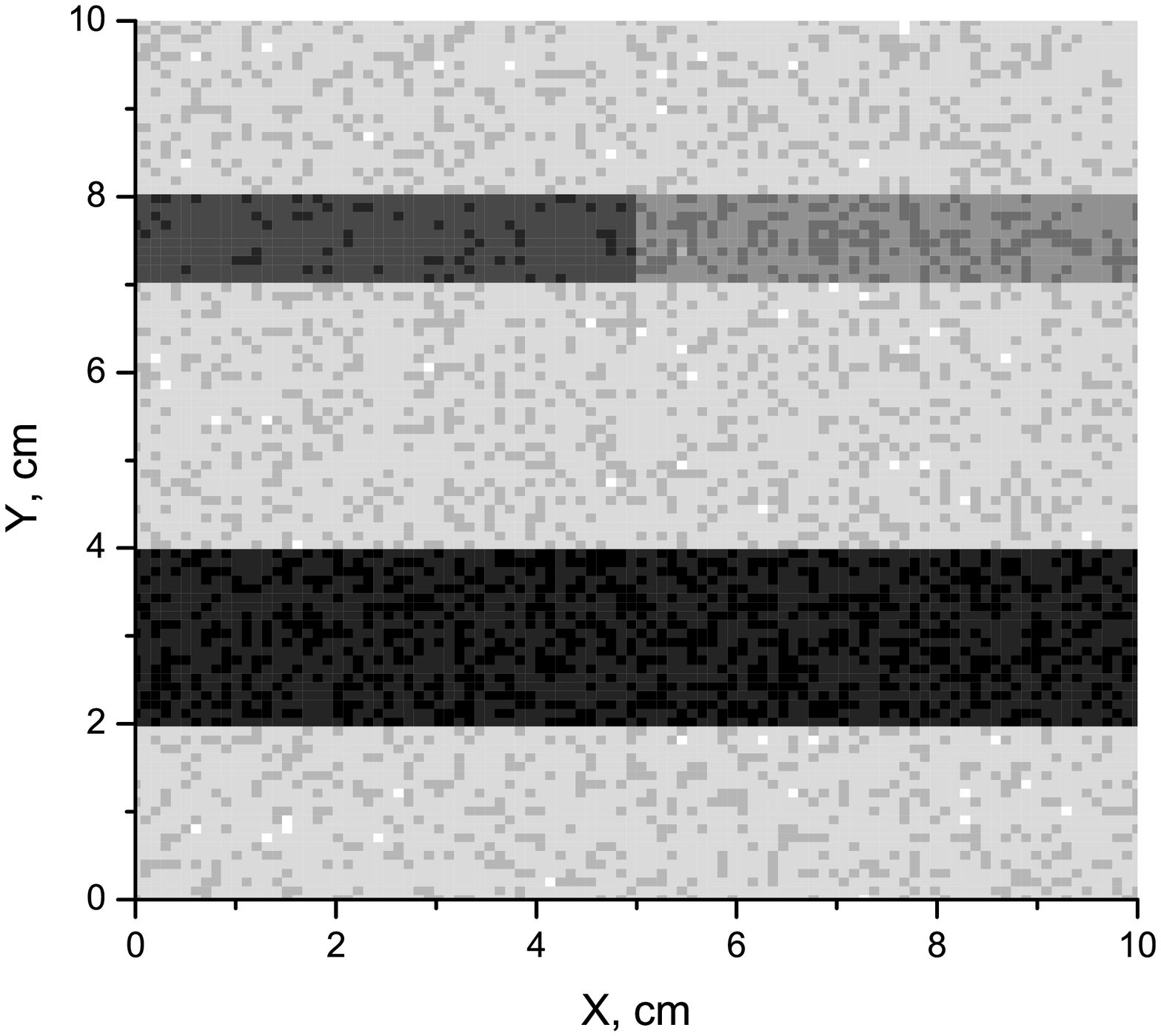}
\includegraphics[width=0.4\textwidth]{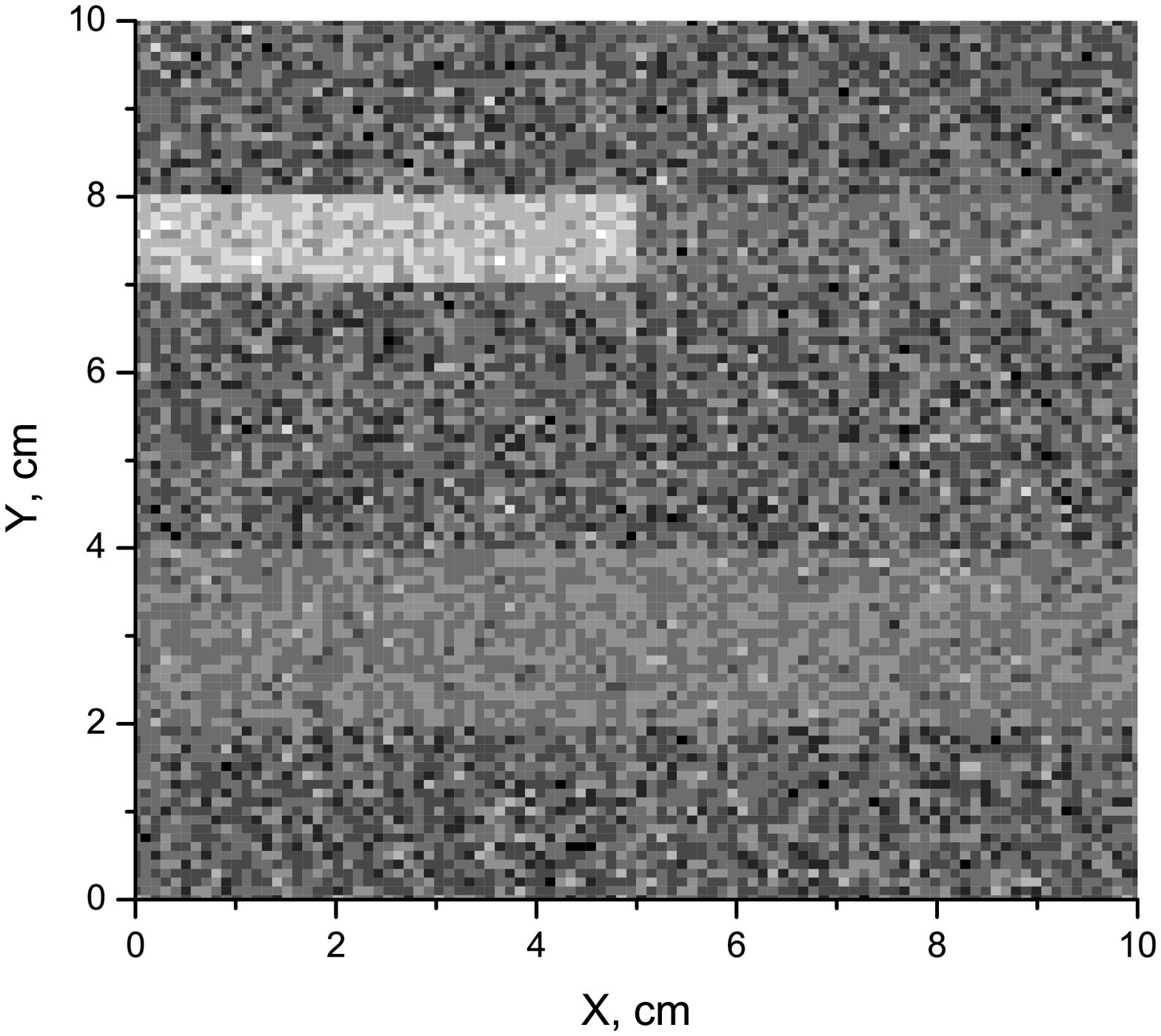}
\caption{\label{fig3b} X--ray image with $E_x^{max} = 35\,\mbox{keV}$ (left) and a subtracted `30\,keV' minus `35\,keV' image (right).}
}
\end{figure}

A run of simulation on the model reveals applicability of Compton x--ray source for angiography as is presented in Fig.\ref{fig3b} (left). The model input data was: x--ray energy ranges 23.3--35\,keV; number of quanta in the range $2\times 10^7$ (total number over the spectrum $2\times 10^7$); the sensor mesh $100\times 100$ pixels (1\,mm$^2$ pixels on $10\times 10\,\mbox{cm}^2$ tissue); the surface densities of tissue's substances are as follows (in g/cm$^2$): muscle 5, bone 0.5, blood 0.5, iodine 0.0125. The muscle component is uniform density over the tissue, the bone--like substance is added at the bottom ($2< y < 4$\,cm), the blood one at the top of the tissue ($7< y < 8$\,cm), iodine is placed in the left half of the blood tape ($x<5$\,cm).

To compare the left half of `35\,keV' image in Fig.\ref{fig3b} (blood with the contrast agent) with the right half (no contrast agent), one can detect location of the iodine. Total flux density for the simulated case is $4\times 10^5$ quanta per cm$^2$.

\subsection{Advanced method of Compton angiography}
Specific properties of the Compton x--ray radiation spectrum --  steep high-energy cutoff with the maximum near it, and strong dependence of the maximal energy upon the energy of electrons, -- enable us to propose an advanced method of Compton angiography.

The method consists in following. Dissimilar to the conventional angiography where location of the contrast agent is revealed as a result of subtracting the image without the agent from that with it, the both images are taken with presence of the contrast agent but one of them at the maximum energy of the spectrum tuned just below the K--edge of the agent, another -- above it. Difference between these images will display the contrast agent location.

Simulations of the method have validated it. In Fig.\ref{fig3b} (right panel) a subtracted image is presented, the image of the described above tissue at the maximum energy of Compton x--rays of 35\,keV (left panel) was subtracted from the image with the maximum energy 30\,keV. Since opaqueness of the agent increases with increasing the energy above K--edge while the other components become more transparent, the location of agent in the difference image is lighter than the others (see Fig.\ref{fig3b}, right panel).

Thus, the advanced method is capable to reveal the location of the contrast agent much faster than the conventional Compton angiography method with about the same flux of x--ray quanta. Speed up of the imaging is due to the fact that injection of the contrast agent into the blood vessel will take much more time than switching of the accelerator energy (we believe it will take a fraction of second).

\subsection{`Proof--of--principle' experiment proposal for LUCX}

The `proof--of--principle' experiment must verify the basic suggestion for Compton x--ray source: opaqueness of a contrast agent increases with increasing the maximal energy of spectra above the K-edge of contrast agent. Also ability of Compton sources to gather sufficient statistics revealing the location of the contrast agent should be proved.

Such an experiment is proposed to conduct at LUCX facility \cite{fukuda13,sakaue13}.
Since the energy of Compton x--ray photons in this accelerator with YAG laser (1\,eV) can not overlap the K--line of iodine, we propose to use bromine as a contrast agent with K-edge at 13.4737\,keV, see \cite{nist}.

\begin{figure}[h]
\centering{
\includegraphics[width=12pc]{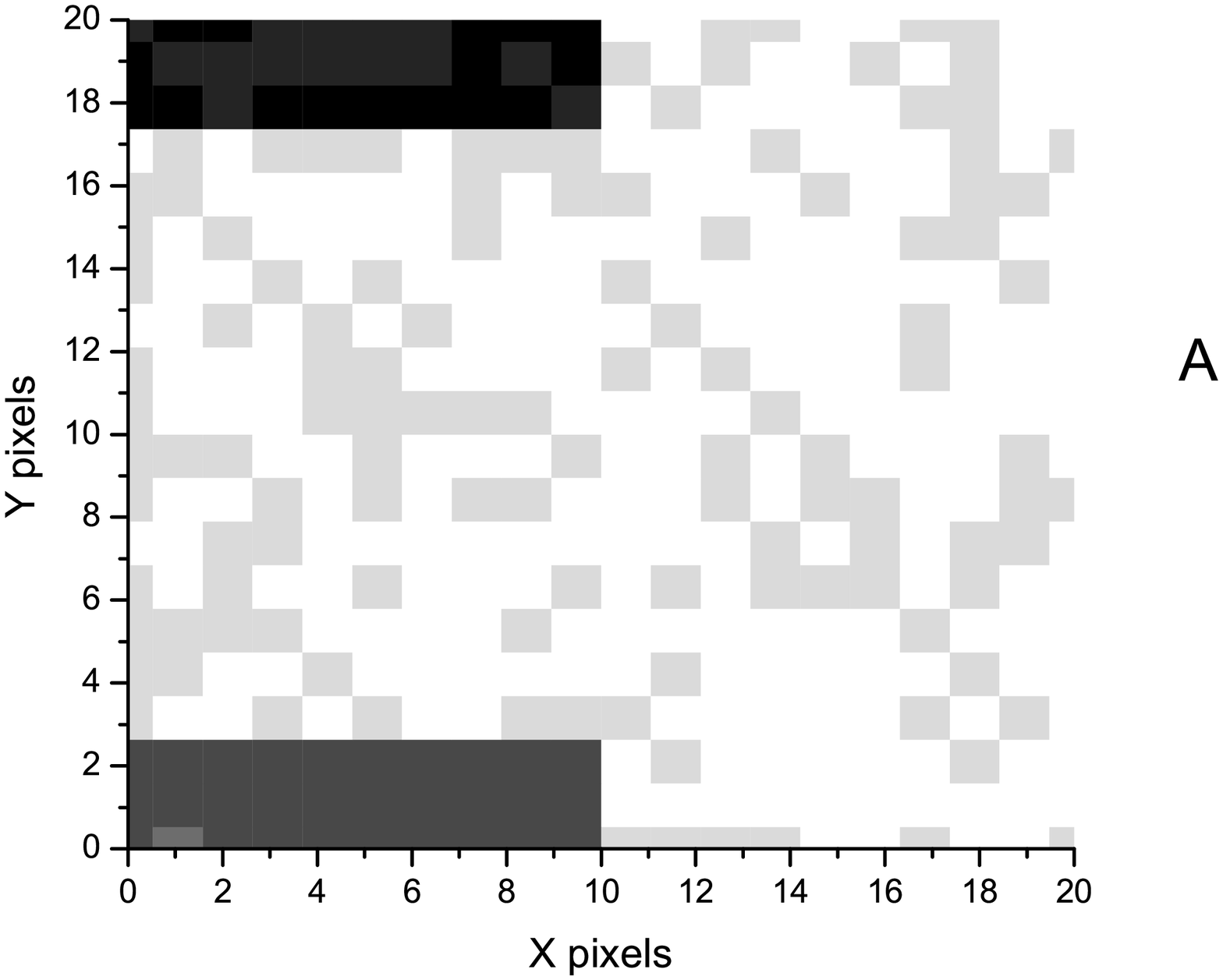}
\includegraphics[width=12pc]{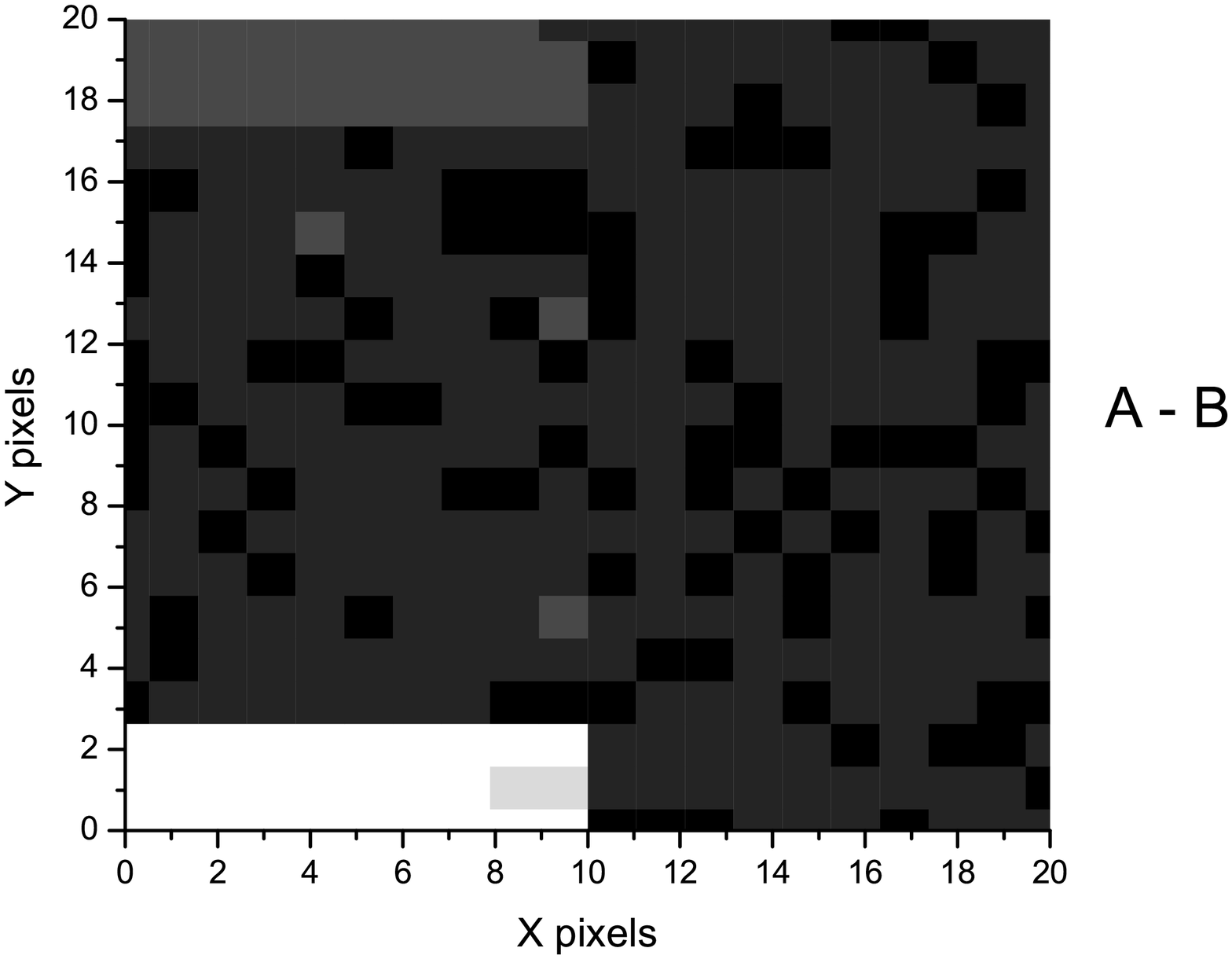}
\includegraphics[width=12pc]{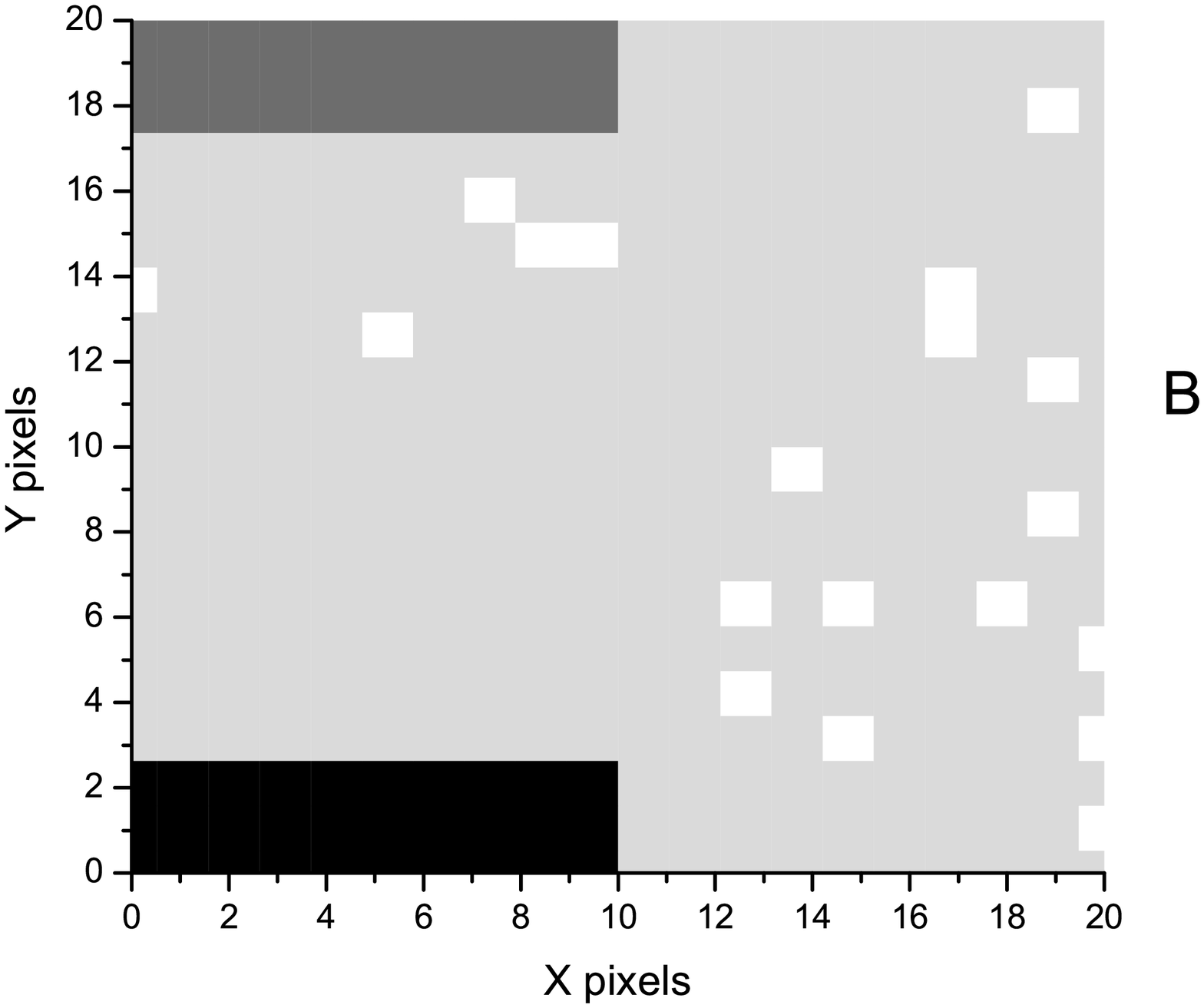}
\caption{\label{figProoP} `13\,keV' image (A, left),  `16\,keV' image (B, right) and the  subtracted  density (A-B, centre).}
}
\end{figure}

We have simulated a `proof--of--principle' experiment for LUCX. Results of the simulation is depicted in Fig.\ref{figProoP}. The sample of $20\times 20$ cells was irradiated by $2\times 10^6$ quanta of each energy, $E_x^{(max)}=13,16\,\mbox{keV}$ (the images `A' and `B', respectively). The Compton spectrum was collimated to $1/\gamma $ opening angle. The sample thickness (muscle skeletal) was 0.5\,g/cm$^2$ with additional 0.15\,g/cm$^2$ in top left corner, bromine thickness was 10\,mg/cm$^2$ (bottom left corner).

As is follows from the images, one cannot tell from single run (A and B images) which of the shadows induced by bromine or by an additional thickness. The bromine shadow is thinner when the spectrum does not overlap the K--peak, `A' image. The subtracted image, `A-B' clearly indicates which of shadows belong to the additional thickness or to the bromine -- the contrast agent is more transparent for lower energy inverse to the regular behavior.

\section{Summary and conclusion}
X--ray radiation generated by the Compton sources has the spectrum with the steep high--energy cutoff, which is independent on the collimation opening angle.
The specific shape of the spectrum with tunable maximum is able to substantially reduce radiation load for the angiography procedure.

We propose a novel angiography procedure consisting in deriving a subtracted image from the two images having been made with different maximum energy of the Compton spectrum, one with the energy below the K--edge of a contrast agent, another -- with the maximum energy overlapped it.

The carried out simulations proved the suggestion of advantage the Compton subtracted scheme for angiography. Also we proposed a `proof--of--principle' experiment that can be conducted at LUCX facility of KEK employing bromine as a radiocontrast agent.

\ack
Authors wishing to acknowledge Prof. A.~Dovbnya, Drs. P.~Gladkikh and V.~Skomorokhov for their assistance and helpful discussions.
Work is supported by the Photon and Quantum Basic Research Coordinated Development Program by the Ministry of Education, Culture, Sports, Science and Technology
`Fundamental Technology Development for High Brightness X-ray Source and the Imaging by Compact Accelerator.'

\end{document}